\documentclass[10pt]{article}
\usepackage{mathrsfs}
\usepackage{amsmath,amsfonts,amssymb,mathtools}
\usepackage{cite}
\usepackage{dsfont}
\usepackage{graphicx}
\usepackage{hyperref}
\usepackage{verbatim}
\usepackage{slashed}
\usepackage[all]{xy}
\usepackage{xcolor}
\usepackage{subfig}
\usepackage{tikz}
\usepackage[T1]{fontenc}
\usepackage{lmodern}
\usepackage{stackengine} %per \ocirc 

\hypersetup{
  colorlinks,
  citecolor=violet,
  linkcolor=blue,
  urlcolor=blue}

\csname @addtoreset\endcsname{equation}{section}
\DeclareMathAlphabet\mathbfcal{OMS}{cmsy}{b}{n}

\newcommand{\beq}{\begin{equation}}
\newcommand{\eeq}{\end{equation}}
\newcommand{\bea}{\begin{eqnarray}}
\newcommand{\eea}{\end{eqnarray}}
\newcommand{\ba}{\begin{array}}
\newcommand{\ea}{\end{array}}
\newcommand{\bit}{\begin{itemize}}
\newcommand{\eit}{\end{itemize}}
\newcommand{\nn}{\nonumber}

\newcommand{\mezzo}{\frac{1}{2}}
\newcommand{\complesso}{{\ \hbox{{\rm I}\kern-.6em\hbox{\bf C}}}}
\newcommand{\reale}{{\hbox{{\rm I}\kern-.2em\hbox{\rm R}}}}
\newcommand{\uno}{ \,  \raisebox{+0.14em}{{\hbox{{\rm \scriptsize ]}} \raisebox{-0.2em}{\kern-.8em\hbox{1}}}} \, }  %  operatore identit\`a

\renewcommand{\a}{\alpha}

\newcommand{\g}{\gamma}

\newcommand{\G}{\Gamma}

\newcommand{\D}{\Delta}

\newcommand{\Er}{{\mathbfcal{E}}}

\renewcommand{\k}{\kappa}

\renewcommand{\L}{\Lambda}

\newcommand{\m}{\mu}

\newcommand{\n}{\nu}
\renewcommand{\r}{\rho}
\newcommand{\s}{\sigma}

\begin{document}

%\begin{comment}

\begin{titlepage}

\begin{flushright}
$LIFT$--2--2.22
\end{flushright}

\vspace{1.5cm}

\begin{center}
\renewcommand{\thefootnote}{\fnsymbol{footnote}}
{\huge \bf Removal of conical singularities from rotating\\ \vskip 5mm  C-metrics and dual CFT entropy
}
\vskip 31mm
\large {Marco Astorino$^{a}$\footnote{marco.astorino@gmail.com}}\\

\renewcommand{\thefootnote}{\arabic{footnote}}
\setcounter{footnote}{0}
\vskip 10mm
{\small \textit{$^{a}$Laboratorio Italiano di Fisica Teorica (LIFT)\\
Via Archimede 20, I-20129 Milano, Italy}\\
} \vspace{0.2 cm}
%{\small \textit{$^{a}$Istituto Nazionale di Fisica Nucleare (INFN), Sezione di Milano \\
%Via Celoria 16, I-20133 Milano, Italy}\\
%} \vspace{0.2 cm}
%{\small \textit{$^{b}$Universit\`a degli Studi di Milano}} \\
%{\small {\it Via Celoria 16, I-20133 Milano, Italy}\\
%}
\end{center}
\vspace{5 cm}

\begin{center}
{\bf Abstract}
\end{center}
{We show how to remove from the rotating C-metric spacetime, which describes accelerating Kerr black holes, both conical singularities. This can be done by embedding the metric into a swirling gravitational universe, through a proper Ehlers transformation. The spin-spin interaction between the external rotational background and the black hole provides the source of the acceleration without the need of rods or strings. \\
The physical properties and entropy of the new solution are studied using near horizon and dual conformal techniques of the Kerr/CFT correspondence.\\
The charged case is also analysed: Accelerating Reissner-Nordstrom and Accelerating Kerr-Newman space-times embedded in a swirling universe are also generated.}

\end{titlepage}

\addtocounter{page}{1}

\newpage

%\tableofcontents
%\newpage

\section{Introduction}

Rotating C-metric are well known in General Relativity to describe accelerating and rotating black holes of the Kerr family. Unfortunately usually these vacuum solutions are plagued by two asymmetrical and unremovable conical singularities on the azimuthal  symmetry axis. From the mathematical point of view these singularities manifest themself as delta-like divergences in the curvature, i.e. in the Riemann tensor. On the other hand, from a physical point of view, these singularities sometimes are interpreted as strings or struts, attached to the event horizon poles, which respectively pull or push the black hole. However the strings are of unbounded length because they extend to infinity; while rods violate all the reasonable energy conditions for matter: basically they have to be made by repulsive matter. Therefore the removal of these singularities from the accelerating black hole spacetimes is a significant task, also because it can provide a physical reasonable reason for their acceleration.  \\
In the literature there have been few intents to remove these conical defects from the C-metric, the most known due to Ernst for the static case \cite{ernst-remove}, or see \cite{marcoa-pair} for the stationary case. The main idea behind those approaches was to embed the accelerating and charged black holes in an external electromagnetic field through a Harrison transformation \cite{ernst-magnetic}. The acceleration in that case is provided by the interaction by the external electromagnetic field of the background, the Melvin universe, and the intrinsic electromagnetic charges of the black hole. Thanks to this interplay the metric remains completely regular outside the event horizon. Some attempts have been pursued also using a gravitational multipolar background, also in this case the pioneering results were derived by Ernst \cite{ernst-generalized-c}. Some further generalisations including accelerating charged and rotating multi black hole configurations were given recently in \cite{marcoa-binary}, \cite{charged-binary}, \cite{multipolar-acc}. The external multipolar approach is very flexible because it furnishes an arbitrary number of external parameters to regularise basically any number of axial conicities, which is useful in multiple collinear configurations. This model is also rather practical because the external multipolar expansion can model a large spectrum of axisymmetric and stationary sources, for instance disk or ring galaxies \cite{deCastro}.
Furthermore the fact that black holes do not need to carry intrinsic electromagnetic charge in order to be regularised is more consistent with actual phenomenological observations, as black holes are believed to be fundamentally neutral in nature. However the external multipolar expansion brings curvature singularities outside the black hole horizon, even though the region where the curvature is unbounded is located asymptotically far away from the black hole and it corresponds to the presence of the source of the external gravitational field, therefore these curvature divergences arise quite naturally. Moreover in the case of the C-metrics this region of spacetime is screened by the Rindler horizon, typical of accelerating  solutions, hence it would not be physically accessible by an observer which lives in the domain of outer communication. Nevertheless it would be even better if we can find a mechanism able to provide a complete regularisation for accelerating metrics in pure vacuum Einstein gravity and without to worry about matter sources.\\
The objective of this paper is to address this problem: provide a physical motivation for the black hole acceleration described by the C-metrics in pure vacuum general relativity, which is mathematically coherent with the basic assumption of the theory, such as the manifold has to be regular outside the event horizon, and also phenomenological viable. Our proposal consists in taking advantage of a sort of spin-spin interaction between the accelerating black hole and the gravitational background.\footnote{This idea is not completely new, it was tested in binary black hole configurations too. However it is well known that this approach is not sufficient in the case of two rotating black holes. The double Kerr metric remains with conical singularity in case it describes two proper black holes, that is when it is not in the hyper-extremal case.} Thus we need to consider an accelerating and stationary black hole and also a rotating universe in which the collapsed star is immersed. At this purpose the results of \cite{swirling} are particularly useful because there it is explained how to embed an arbitrary stationary and axisymmetric seed solution into a swirling universe, a gravitational whirlpool where the frame dragging of the background universe deforms and interacts with the black hole. \\
Hence in section \ref{sec:generation} we generate the new solution thanks to a kind of Ehlers transformation. Starting with the rotating C-metric as a seed, we embed the accelerating Kerr black hole into a rotating universe. In section \ref{sec:properties} we study some physical and geometrical properties of the metric and we show that both the conical singularities of the rotating C-metric can be simultaneously removed from the spacetime. In section \ref{sec:cft-thermo}, thanks to some tools provided from the near horizon analysis and conformal field theories, we study the entropy of the extreme black hole. Finally in section \ref{sec:charged} we extend the generating transformation to the presence of an Maxwell electromagnetic field, thus we are able to discuss accelerating Kerr-Newman and Reissner-Nordstrom black holes in the swirling background.\\

\section{Embed Accelerating Kerr black Holes in a Swirling Universe}
\label{sec:generation}

To embed a given arbitrary stationary and axisymmetric seed solution into a rotating background, such as the swirling universe studied in \cite{swirling} we make use of the results presented therein. The procedure is valid in the realm of general relativity, which is governed by vacuum Einstein equations $R_{\mu\nu}=0$. In this setting the most generic metric, possessing two commuting Killing vector compatible with the above symmetries ($\partial_t,\partial_\varphi$), can be written in the Lewis-Weyl-Papapetrou form

\beq
\label{LWP-metric}
{ds}^2 = -f ( d\varphi - \omega dt)^2 + f^{-1} \bigl[ \rho^2 dt^2 - e^{2\gamma}  \bigl( {d \rho}^2 + {d z}^2 \bigr) \bigr] \ \ ,
\eeq
where the three functions $f,\omega, \gamma$ in the metric are depending only on the non-Killing coordinates $(\rho,z$). 
The vacuum Einstein's equations are equivalent to the Ernst's equations
\beq
\label{ee-ernst}  \bigl( \Er + \Er^* \bigr) \nabla^2 \Er   =   2 \, \vec{\nabla} \Er  \cdot \vec{\nabla} \Er  \ \ ,
\eeq
for a complex scalar function called Ernst potential, $\Er(\rho,z) \coloneqq f(\rho,z) + i h(\rho,z) $, where
\beq
\label{h}
\vec{\nabla}   h \coloneqq -
\frac{f^2}{\rho} \vec{e}_\varphi \times \vec{\nabla} \omega \ \ .
\eeq
A great advantage of the Ernst equations consist in some Lie-point symmetry they enclose, such as the Ehlers transformation
\beq
\label{ehlers-transf}
  \Er \to \Er' = \frac{\Er}{1+i\jmath\Er} \ .
\eeq
This one real parameter transformation, labelled by $ \jmath $, is usually known in the literature because it adds the NUT parameter to a given seed metric \cite{reina-treves}, \cite{enhanced}. On the other hand we would like to use it in an alternative way, basically combining it with a couple of discrete symmetries, as done in \cite{swirling}, in order to add a rotating background to a given seed solution, $ds_0^2$.\\  
The $\gamma(\rho,z)$ function of the LWP metric (\ref{LWP-metric}) not only remains decoupled from the equations for $f$ and $\omega$, but its equations are automatically satisfied under the Ehlers transformation, for any given initial seed $\gamma_0$, so it keeps unchanged under the transformation (\ref{ehlers-transf}).\\
Since we want to embed an accelerating and rotating Kerr black hole into a spinning universe we have to start, as a seed, with the following rotating C-metric\footnote{Here $(r,x=\cos\theta)$ are the standard spherical coordinates.}, in terms of the general metric form  (\ref{LWP-metric})
\bea
    \label{f0}  f_0(r,x) &=& \frac{a^2(1-x^2)^2\Delta_r(r) - (a^2+r^2)^2 \Delta_x(x)}{(1-Arx)^2(r^2+a^2x^2)} \ \ ,\\ 
      \omega_0(r,x) &=& \frac{a(1-x^2)\Delta_r(r)-a(a^2+r^2)\Delta_x(x)}{(a^2+r^2)^2 \Delta_x(x)-a^2(1-x^2)^2\Delta_r(r)} \ \ , \\
      \gamma_0(r,x) &=& \frac{1}{2} \log \left[ \frac{(a^2+r^2)^2 \Delta_x(x)-a^2(1-x^2)^2\Delta_r(r)}{(1-Arx)^4} \right] \\
   \label{rho0}   \rho^2(r,x) &=& \frac{\Delta_x(x) \Delta_r(r)}{(1-Arx)^4}  \ \ , \\
\label{Dr}    \Delta_r(r) &=& (1-A^2r^2)(r^2-2mr+a^2) \quad , \\
 \label{Dx}   \Delta_x(x) &=& (1-x^2) (1-2Amx+a^2A^2x^2) \ \ . 
\eea
Thanks to (\ref{h}) we can deduce the imaginary part of the Ernst seed potential $h_0(r,x)$, up to an arbitrary constant which affect the angular speed of an asymptotic observer and we dismiss without loss of generality
\beq
         h_0(r,x) = \frac{2am\left\{r^2 \left[-3x+x^3+Ar(1+x^2) \right]-a^2\left[x+x^3+A(r-3rx^2)\right]\right\}}{(1-Arx)(r^2+a^2x^2)} \ \ .
\eeq
Thus combining this latter with (\ref{f0}) we have the complex Ernst potential $ \Er_0(r,x) = f_0(r,x) + i h_0(r,x) $ associated to the seed metric (\ref{f0})-(\ref{Dx}). To embed the rotating C-metric into a swirling universe we apply the Ehlers transformation (\ref{ehlers-transf}), obtaining a new Ernst potential
\beq \label{E1}
      \Er = \frac{\Er_0}{1+i\jmath\Er_0} \ ,
      \eeq
which represents a new solution. The metric functions for the new space-time can be obtained by the definition of the Ernst potential: $f$ from the real part of (\ref{E1}), while $\omega$ from the imaginary part of $\Er$ and (\ref{h}). The explicit expression for $\omega(r,x)$ is given in appendix \ref{app-w}\footnote{Moreover a Mathematica notebook containing the full solution can be find in between the source arXiv files.}. The resulting metric can be cast in a similar form with respect to the Kerr-Melvin spacetime
\beq \label{swirling-sol}
        ds^2 = -\frac{f_0(r,x)}{|\Lambda(r,x)|^2} \left( \Delta_\varphi d\varphi - \omega(r,x) dt \right)^2 + \frac{|\Lambda(r,x)|^2}{f_0(r,x)} \left[ \rho^2(r,x) dt^2 - e^{2\gamma_0(r,x)}  \left( \frac{{d r}^2}{\Delta_r(r)} + \frac{{d x}^2}{\Delta_x(x)} \right) \right] \ \ .
\eeq
where 
\beq
       \Lambda(r,x) := 1 + i \jmath \Er_0(r,x) = 1 + i \jmath f_0(r,x) - \jmath h_0(r,x)
\eeq
and $\Delta_\varphi$ is a constant that can be useful to regularise the metric without rescaling the periodicity of the azimuthal angle range which is $[0,2\pi]$, as usual. Because of the form of the metric (\ref{swirling-sol}) the causal structure of the spacetime does not differ much with respect to the accelerating Kerr black hole. In fact we have an inner horizon and event horizon located at the zeros of $\Delta_r(r)$ as in Kerr spacetime
\beq
              r_\pm = m\pm \sqrt{m^2-a^2} \ \ .
\eeq
Also it is easy to check absence of closed time-like curves\footnote{We recall that in order to have closed time-like curves the $g_{\varphi\varphi}$ element of the metric must take negative values.} because the $g_{\varphi\varphi}$ element of the metric is just dilatation, by the positive factor $1/|\Lambda(r,x)|^2$ of the seed accelerating Kerr element, therefore it preserves the positive signature of the seed element. 
Moreover the curvature invariants, such as the Kretschmann scalar, single out the usual ring curvature singularity inside the black hole horizon. Of course also the typical conformal factor of the accelerating metrics remains. It determinates the acceleration horizon, and the location of conformal infinity, as in the case for $\jmath=0$.\\  
We note that even though the new solution is generated by an Ehlers transformation, the function $\omega(r,\theta)$ is regular on the symmetry axis, 
\begin{equation}
\lim_{\theta\to0} \ \frac{g_{t \varphi}}{g_{tt}}
= \lim_{\theta\to\pi} \ \frac{g_{t\varphi}}{g_{tt}}
= 0 \,,
\end{equation}
hence the new metric is not plagued by Misner strings or NUT defects. That is because we are not using the Ehlers transformations as done in \cite{reina-treves} or \cite{enhanced}, but as proposed in \cite{swirling}\footnote{The two Ehlers transformations are related by a couple of discrete conjugate transformations. Or alternatively one can think to use the same Ehlers transformation but on two different variations of the LWP metric. More details in \cite{swirling}.
}. \\
Note also that, similarly to the Kerr-Melvin, Kerr-AdS or the accelerating Kerr solution the angular velocity on the axis of symmetry $\Omega_z=g_{t\varphi}/g_{\varphi\varphi}|_{x\pm1}$ is not a constant, as in asymptotic flat spacetimes.\\

The Ehlers transformation (\ref{E1}) immerses the accelerating Kerr black hole into an axial symmetric rotational background as described in \cite{swirling}. The frame dragging of the background can have several origins, for instance, it is thought to be caused by two asymptotic far away  counter rotational sources on the $z$-axis, which however do not contribute to the energy-momentum tensor. It could model, with a certain degree of approximation, the gravitational setting that is generated by the interplay of two counter rotating black holes or galaxies.

This method can be directly extended in the presence of any Maxwell electromagnetic field compatible with the symmetries of the metric, just using the generalised Ehlers transformation instead of the eq. (\ref{ehlers-transf}), as explained in \cite{enhanced}. Further generalisations to the presence of minimally or conformally coupled scalar fields are also possible, thanks to the results of \cite{marcoa-embedding},\cite{marcoa-stationary}. Note that the seed metric could represent any space-time, it does not necessarily have to be a black hole, it could be a wormhole or anything respecting axisymmetry and stationarity.\\

\section{Geometrical properties of the Black Hole}
\label{sec:properties}

The fundamental point of our work consists in the analysis of the conical singularities on the polar axis, to clarify if the interaction between the spin of the black hole and the rotation of the background can support the C-metric acceleration. \\
An angular defect can be detected when computing the ratio between the perimeter and the radius of a small circle around the $z$-axis on the north and south poles, for $x=\pm1$. When these quantities are different from $2\pi$ the metric is plagued with conical singularities. In the coordinate we are using these ratios are
\beq \label{cony-1}
      \lim_{x \to -1} \int_0^{2\pi} \frac{d\varphi}{1-x^2} \sqrt{\frac{g_{\varphi\varphi}}{g_{xx}}} = 2\pi \Delta_\varphi \frac{1+a^2A^2+2mA}{(1+4a \jmath m)^2} 
\eeq
and 
\beq \label{cony+1}
      \lim_{x \to +1} \int_0^{2\pi} \frac{d\varphi}{1-x^2} \sqrt{\frac{g_{\varphi\varphi}}{g_{xx}}} = 2\pi \Delta_\varphi \frac{1+a^2A^2-2mA}{(1-4a \jmath m)^2}  \ \ .
\eeq
Remarkably it is possible to impose that both (\ref{cony-1}) and (\ref{cony+1}) take the standard value for the azimuthal angle, i.e. $2\pi$,  by fixing one parameter of the solution and the gauge constant $\D_\varphi$; for instance as%\footnote{Note that the continuous, and regular, limit to the non swirling solution, i.e. kerr ($\Delta \varphi \to 1$, $\jmath \to 0$), can be obtained chosing the $+$ sign in (\ref{c})-(\ref{Deltaphi}) for $a>0$, while viceversa with the negative sign for $a<0$.}
\bea \label{c}
      \jmath &=& \frac{a + a^3A^2 \pm \sqrt{a^2[(1+a^2A^2)^2-4A^2m^2]}}{8a^2Am^2} \\
   \label{Deltaphi}   \Delta_\varphi &=& \frac{a + a^3A^2 \pm \sqrt{a^2[(1+a^2A^2)^2-4A^2m^2]}}{2aA^2m^2}
\eea
These constraints (\ref{c})-(\ref{Deltaphi}) makes the solution free from conical singularities, hence the spacetime becomes completely regular outside the event horizon. This means that instead of the usual picture where string or strut have to be postulated to justify the acceleration of the black holes, we can interpret the acceleration as the ``force'' exerted by the spin-spin interaction between the angular momentum (for unit mass) of the black hole $a$ and the rotational parameter $\jmath$ of the stationary gravitational background. The compatibility of the regularity constraints  (\ref{cony-1}) and (\ref{cony+1}) in the non-relativistic limit gives an approximation for the intensity of the spin-spin Newtonian force exerts on the two black sources to maintain equilibrium. In fact imposing an equality between expressions (\ref{cony-1}) and (\ref{cony+1}), for small values of the acceleration parameter $A$ and the rotational parameter $\jmath$, we can extract a first order approximation for the ``Newtonian force'' provided by the spin-spin interaction:
\beq
         m A \approx 4a \jmath \ .
\eeq

The metric with the constraints (\ref{c})-(\ref{Deltaphi}) has a continuous and smooth limit in case of null accelerations, $A \to 0$, to the usual Kerr black hole. Indeed in case of zero acceleration also the external swirling gravitational background have to vanish, $\jmath = 0$, if the regularity constraints are implemented\footnote{Of course if we do not care about having a conical metric the rotational background can survive also in absence of acceleration. In that case the metric describes a Kerr black hole in a rotational external backround.\\ This again resemble the behaviour of the Melvin magnetic background: when the intrinsic magnetic charge of the black hole is present, the metric has a non removable conical singularity, unless acceleration is introduced \cite{ernst-remove}.}. Note that in order this limit is well behaved in (\ref{c})-(\ref{Deltaphi}) the positive signs have to be associated with negative values of the angular momentum $a$ and viceversa. Similarly in the case one wants to switch off the external gravitational field, $\jmath\to0$ while keeping the solution free from conicity, thanks to constraints (\ref{c})-(\ref{Deltaphi}), the only possibility is vanishing the acceleration parameter $A$. It's easy to see how the constraints (\ref{c})-(\ref{Deltaphi}) diverges for $a\to0$ unless also $A=0$, therefore the only black hole regular solution of this family without intrinsic angular momentum is the non accelerating Schwarzschild black hole embedded in the swirling universe, as discussed in \cite{swirling}.\\

We stress the fact that the spin-spin interaction is the fundamental physical reason that justify the acceleration of the black hole and that other different intrinsic properties of the metric, such as the charge or the mass cannot provide a motivation for this effect. From a mathematical perspective it means that the space-time remains with at least a conical singularity when $A\neq0$. In fact, as we have seen in (\ref{c})-(\ref{Deltaphi}), we cannot regularise the accelerating spacetime by means of the only mass parameter, i.e. without the presence of $a$. Moreover as we will discuss in section \ref{sec:charged} the presence of the electric or magnetic charge alone (without the presence of intrinsic angular momentum parametrised by $a$) does not guarantee the removal of the nodal singularities from both the north and the south poles.\\

The event horizon area can be computed by
\beq
\label{area}
        \mathcal{A_+} = \int_0^{2\pi}  \int_{-1}^{1} \sqrt{g_{xx} g_{\varphi\varphi}} dx  d\varphi  =   \frac{8\pi  \Delta_\varphi m r_+}{1-A^2r_+^2} \ . 
\eeq

The Killing vector generator of the black hole horizon is 
\beq
        \chi^\mu = \partial_t + \Omega_H \partial_\varphi
\eeq
where
\beq \label{OmegaH}
             \Omega_H = - \frac{g_{t\varphi}}{g_{\varphi\varphi}} = \frac{-2\jmath^2m+A^4(2m-r_+)[-1+8\jmath^2 m^2(3m-2r_+)r_+]+2A^2\jmath^2m(-4m^2-2mr_++r_+^2)}{2aA^4m\Delta_\varphi} \ \ .
\eeq
The temperature $T_H$ of the black hole is given in term of the surface gravity on the event horizon $\k_s$ by
\beq
      T_H =\frac{\k_s}{2\pi} = \sqrt{ -\mezzo  \nabla_\mu \chi_\nu \nabla^\nu \chi^\mu } \ \bigg|_{r=r_+} =  \frac{r_+-m}{2mr_+} \ \frac{1-A^2r_+^2}{2\pi} \ \ .          
\eeq

Hence nor the area of the black hole, nor its temperature are directly affected by the presence of the external gravitational frame dragging. In fact they have the same form of the $\jmath=0$ case, as occurs when the accelerating Kerr spacetime is distorted by the external Melvin electromagnetic field \cite{magnetised-kerr-cft}. \\
As discussed in detail in \cite{swirling} this family of solutions does not belong to the type D of the Petrov classification, apart the pure background subcase (where all the parameters are null except $\jmath$). This is once more in analogy to the magnetised black holes.\\
Note that all the results here are valid both in the general case and in the regularised case, where the parameters are constrained by (\ref{c})-(\ref{Deltaphi}) to avoid the presence of string or strut. Also the correspondence with the conformal field theory, as described below, holds in the non accelerating limit. \\

\section{Near horizon geometry, extremal Kerr/CFT  and dual microscopic Entropy}
\label{sec:cft-thermo}

We are interested in studying the entropy of the black hole through the tools furnished by the Kerr/CFT correspondence, which provide an alternative way to approach the problem. It relies on the correspondence between the geometry of the black hole near the horizon and a conformal field theory model that is defined on the boundary of the near horizon metric. This approach has the advantage to avoid issues related to the non conventional asymptotic behaviour of the metric, because it focuses mainly on the vicinity of the black hole horizon. It has already proved successful for magnetised and accelerating black holes \cite{magnetised-kerr-cft}, \cite{magnetised-RN-cft}, \cite{c-cft} and also for multi black hole configurations \cite{enhanced}. \\
We will treat in particular the case where the inner and the outer horizon coincides, i.e. $r_\pm=r_e$, because at extremality the black hole near horizon geometry has some enhanced symmetries properties, as clarified in \cite{lucietti-kunduri}. In fact if we define some new dimensionless coordinates adapted to the near horizon description, such as
\beq
      r(\tilde{r}):= \tilde{r}_e+\lambda r_0 \tilde{r} \quad , \qquad t(\tilde{t}):=\frac{r_0}{\lambda} \tilde{t} \quad , \qquad \varphi(\tilde{t},\tilde{\varphi}):=\tilde{\varphi}+\Omega_H^{ext} \frac{r_0}{\lambda} \tilde{t} \quad ,
\eeq
where $\Omega_H^{ext}$ represents the extreme limit of the angular velocity (\ref{OmegaH}), while $r_0$ is a constant introduced for dimensional reasons. We observe that the near horizon extreme black hole metric can be cast into a warped and twisted product of $AdS_2 \times S^2$, as follows
\beq \label{near-metric}
            d\tilde{s}^2 = \Gamma(x) \left[ -\tilde{r}^2 d\tilde{t}^2 + \frac{d\tilde{r}^2}{\tilde{r}^2} + \alpha^2(x) \frac{dx^2}{1-x^2} + \gamma^2(x) \Big( d \tilde{\varphi}+\kappa \tilde{r} d\tilde{t} \Big)^2 \right] \quad ,
\eeq
where
\begin{align}
\label{near-functions-inizio}
                  \Gamma(x) & = \frac{a^2[1-16a^2 \jmath x+x^2+16a^4 \jmath^2 (1+x^2)]}{(1-a^2A^2)(1-aAx)^2} \quad , \hspace{1.8cm} r_0     = \pm \frac{a \sqrt{2}}{\sqrt{1-a^2A^2}}  \quad , \\  
                  \gamma(x)  & = \pm \frac{2 a^2 \sqrt{1-x^2} \D_\varphi^{ext}}{\G(x)\sqrt{1-a^2A^2}(1-A a x)} \quad ,  \hspace{3.4cm}    \kappa  = \frac{-1+16a^4\jmath^2}{\Delta_\varphi(1-a^2A^2)} \quad , \\
                 \alpha(x) & = \pm \frac{\sqrt{1-a^2A^2}}{1-aAx}   \quad .           
\label{near-functions-fine}
\end{align}
Note that this near horizon geometry differs from the one of Kerr, basically because the metric, at this point, is not considered necessarily regular: we are not considering the constraints (\ref{c})-(\ref{Deltaphi}). In section \ref{sec:meissner} we will show how the near horizon geometry simplifies to the Kerr one, once the metric is regularised by fixing the values of $\jmath$ and $\Delta_\varphi$ which ensure the absence of string or struts and their related energy momentum tensors, i.e. as in eqs. (\ref{c})-(\ref{Deltaphi}).

According to the Kerr/CFT correspondence \cite{strominger-kerr-cft}, \cite{compere-kerr-cft} the asymptotic regions of the near horizon metric (\ref{near-metric})-(\ref{near-functions-fine}) can give some information about the microscopic entropy of the extremal black hole entropy. In fact on that boundary the near horizon fields enjoy a conformal symmetry. More precisely a two dimensional dual conformal field theory is thought to live in the asymptotic region of the near horizon metric. For details about the required boundary condition and its symmetries we refer to \cite{magnetised-kerr-cft} or \cite{c-cft}, where the same notations are used. \\
The left central extension of the Virasoro algebra for the dual CFT can be computed as follows
\beq
            c_L = 3 \k \int_{-1}^{1} \frac{dx}{\sqrt{1-x^2}} \G(x) \a(x) \g(x) = \frac{12a^2(-1+16\jmath^2 a^4)}{(1-a^2A^2)^2} \quad .
\eeq 
Thanks to the Cardy formula 
\beq \label{cardy}
         \mathcal{S}_{\mathcal{CFT}} = \frac{\pi^2}{3} c_L T_L 
\eeq
it is possible to obtain the microscopical entropy associated to the dual model on the boundary. The Hawking temperature $T_H$ of the black hole is null at extremality, thus in the Kerr/CFT approach the Frolov-Thorne temperature is used as left temperature $T_L$ in the Cardy formula to take into account the rotational degrees of freedom. The Frolov-Thorne temperature $T_L$ can be inferred by the following limit
\beq
            T_L := \lim_{r_+ \to r_e} \frac{T_H}{\Omega_H^{ext} - \Omega_H} = - \frac{\Delta_\varphi^{ext}(1-a^2A^2)}{2\pi(1-16\jmath^2 a^4)} = \frac{1}{2\pi \k} \quad . 
\eeq
where $\Delta_\varphi^{ext}$ represents the extremal limit of $\Delta_\varphi$ constant. Finally the entropy of the conformal field theory associated to the extremal near horizon geometry of the black hole (\ref{cardy}) becomes
\beq
        \mathcal{S}_{\mathcal{CFT}} =  \frac{2a^2\pi\Delta_\varphi^{ext}}{1-a^2A^2} = \mathcal{S_{BH}} \quad .
\eeq
Note that this value precisely recovers the standard Bekenstein-Hawking black hole entropy $\mathcal{S_{BH}}$: a quarter of the horizon area, as can be easily seen taking the extremal limit of (\ref{area}). Therefore the Kerr/CFT correspondence reveals to be a consistent tool to extract physical information from the near horizon geometry of the black hole, to map it into the dual CFT and vice versa, at least at extremality. However outside the extremal limit its applicability is less straightforward because, for involved metrics as the one we are considering here, the separability of the Klein-Gordon equation in the vicinity of the black hole is difficult to achieve. Other near horizon analysis techniques, such as the one in \cite{gaston}, may result more effective at this purpose.    \\

\section{Rotational Meissner effect?}
\label{sec:meissner}

Since the similitude between the rotational gravitational background considered in this article and the electromagnetic background of the Melvin universe is strict, we would like to explore a possible behaviour for these extremal black hole analogous to the Meissner effect, in the presence of external electromagnetic field. In fact it is well known that black holes embedded in an external Melvin universe eject the electromagnetic field when their temperature is lowered to zero, i.e. at extremality \cite{magnetised-kerr-cft}, \cite{bicak}. This is in analogy with the standard Meissner effect which consists in  the expulsion of the magnetic field from some superconductive materials when cooled under a certain critical temperature.  \\
If the analogy holds we expect that black holes in an external swirling background considered here might expel the external gravitational field at zero temperature, in practice becoming isometric to the extreme accelerating Kerr one. The easiest way to test this conjecture is looking, at extremality, for an isometry between the near horizon of the (extremal) rotating C-metric and the accelerating Kerr black hole embedded in the swirling background. The near horizon geometry for the latter spacetime is described by eqs. (\ref{near-metric})-(\ref{near-functions-fine}), while near horizon of the first metric can be obtained setting $\jmath=0$:
\begin{align}
\label{ring-functions-inizio}            
                  \mathring{\Gamma}(x) & = \frac{\mathring{a}^2(1+x^2)}{(1-\mathring{a}^2 \mathring{A}^2)(1-\mathring{a}\mathring{A}x)^2} \quad , \hspace{3.3cm} \mathring{\alpha}(x)  = \pm \frac{\sqrt{1-\mathring{a}^2\mathring{A}^2}}{1-\mathring{a}\mathring{A}x}  \quad , \\  
                  \mathring{\gamma}(x)  & = \pm \frac{2 \mathring{a}^2 \sqrt{1-x^2} \mathring{\D}_\varphi^{ext}}{\mathring{\G}(x)\sqrt{1-\mathring{a}^2\mathring{A}^2}(1-\mathring{A} \mathring{a} x)} \quad ,  \hspace{3cm}    \mathring{\kappa}  = \frac{-1}{\mathring{\Delta}_\varphi(1-\mathring{a}^2\mathring{A}^2)} \quad . \\
\label{ring-functions-fine}
\end{align}
In order to distinguish the non-swirling quantities we labelled them with a circle. Then to compare the two near horizon geometries it is convenient to shift both near horizon metrics to the gauge where $\a=1$ and $\mathring{\a}=1$, thanks to the changes of $x$ coordinate 
\beq
        x \longrightarrow \frac{aA+ \hat{x}}{1+aA \hat{x}} \qquad \qquad  , \qquad \qquad  x \longrightarrow \frac{\mathring{a}\mathring{A}+\hat{x}}{1+\mathring{a}\mathring{A} \hat{x}}
\eeq 
The resulting near horizon metric takes the form 
\beq
\label{hats}
          d\hat{s}^2 = \hat{\G}(\hat{x}) \left[ -r^2 dt^2 + \frac{dr^2}{r^2} + \frac{d\hat{x}^2}{1-\hat{x}^2} + \hat{\gamma}^2(\hat{x}) \Big( d\varphi + \hat{\kappa} r dt \Big)^2 \right] \quad ,
\eeq
both in the accelerating and rotating background case and also without the external gravitational field. In the new set of coordinates the near horizon functions, becomes 
\begin{align}
\label{func-ini}
              \hat{\G}(\hat{x}) & = a^2 \frac{(1+aA\hat{x})^2}{(1-a^2A^2)^3} \left\{ 1 + \frac{(aA+\hat{x})^2}{(1+aA\hat{x})^2} -\frac{16 a^2 \jmath (aA+\hat{x})}{1+aA\hat{x}} + 16 a^4 \jmath^2 \left[ 1 + \frac{(aA+\hat{x})^2}{(1+aA\hat{x})^2} \right] \right\} \  ,\\
               \hat{\g}(\hat{x}) & = \frac{2 a^2 \sqrt{1-\hat{x}^2} \D_\varphi}{1-a^2A^2} \ , \\
               \hat{\k} & = - \frac{1-16 a^4 \jmath^2}{\D_\varphi (1-a^2A^2)} \ ,
\label{func-fin}
\end{align}
and 
\begin{align}
\label{func0-ini}
             \mathring{\G}(\hat{x}) & = \frac{\mathring{a}^2 [1+4\mathring{a}\mathring{A}\hat{x}+\hat{x}^2+\mathring{a}^2\mathring{A}^2 (1+\hat{x}^2)]}{(1-\mathring{a}^2\mathring{A}^2)^3} \ ,  \\
    \mathring{\g}(\hat{x}) & = \frac{2 \mathring{a}^2 \sqrt{1-\hat{x}^2} \D_\varphi}{1-\mathring{a}^2\mathring{A}^2} \ , \\
     \hat{\mathring{\k}} & = - \frac{1}{\D_\varphi (1-\mathring{a}^2 \mathring{A}^2)}  \ ,
\label{func0-fin}
\end{align}
It is not difficult to see that the near horizon geometry described by (\ref{hats}), (\ref{func0-ini})-(\ref{func0-fin}) can be mapped in the one determined by (\ref{hats}), (\ref{func-ini})-(\ref{func-fin})
by the following redefinition of the parameters
\beq
         \mathring{a} \to \frac{a(1-16a^4\jmath^2)^{3/2}}{(1-4a^3A\jmath)} \qquad \qquad  , \qquad  \qquad \mathring{A} \to \frac{(A-4a\jmath)(1-4a^3A\jmath)}{(1-16a^4\jmath^2)^{3/2}} \quad ,
\eeq
along with a proper rescaling of the time and azimuthal angle. Therefore the near horizon geometries of the accelerating Kerr with and without the external gravitational background are equivalent. Physically it means that, at zero temperature, the rotational contribution of the background becomes irrelevant near the horizon. This is an analogous process also the magnetised black holes experience \cite{magnetised-kerr-cft}, known as the Meissner effect. It would be interesting to check if this effect survives, outside the near horizon region, in the rotational background field: can the metric (\ref{swirling-sol}) become isometric to the Kerr black hole, at extremality. 

When the regularity constraints (\ref{c})-(\ref{Deltaphi}) are imposed the near horizon geometry becomes even more peculiar. In fact some uniqueness theorems \cite{lucietti-kunduri} ensure that the spacetime near the horizon is even more rigid when the solution is regular (outside the horizon): in the context of general relativity it has to coincide with the one of extreme Kerr, defined by (\ref{hats}) and 
\beq
 \hat{\G}(\hat{x})  = \mathring{a}_0^2 (1 + \hat{x}^2)  \hspace{0.2cm} , \hspace{1.5cm}  
                 \hat{\g}(\hat{x})  = \frac{2 \mathring{a}_0^2 \sqrt{1 - \hat{x}^2}}{\hat{\G}(\hat{x})} \hspace{0.2cm} , \hspace{1.5cm}  \hat{\k}  = -1 \ \ .
\eeq
This correspondence is easily achieved just by rescaling the parameter $a$ as follows
\beq
                  a \rightarrow \frac{\mathring{a}_0}{\sqrt{1+\mathring{a}_0^2 \mathring{A}_0^2}} \ \ .
\eeq

In view of these observations, the fact that the dual conformal picture works well also in this context could be ascribed to the analogy, at extremality, between the standard accelerating Kerr metric and the one embedded into the rotating background. However it should not be considered a totally trivial result because some ingredients of the Kerr/CFT correspondence, such as the left temperature, are obtained also from non extremal quantities. \\

\section{Charged generalisation: Accelerating Reissner-Nordstrom and Accelerating Kerr-Newman black holes in a swirling universe}
\label{sec:charged}

The electromagnetic field can be brought in this picture seamlessly because of the straightforward  generalisation of the Ernst equations (\ref{ee-ernst}) and their symmetries, in the presence of the Maxwell contribution \cite{ernst2}, \cite{enhanced}. Indeed the Einstein-Maxwell equations of motion, 
\bea  \label{field-eq-g}
                        &&   R_{\m\n} -   \frac{R}{2}  g_{\m\n} =   F_{\m\r}F_\n^{\ \r} - \frac{1}{4} g_{\m\n} F_{\r\s} F^{\r\s}  \quad ,   \\
       \label{field-eq-A}                  &&   \partial_\m ( \sqrt{-g} F^{\m\n}) = 0  \ \quad , 
 \eea
for the Lewis-Weyl-Papapetrou metric (\ref{LWP-metric}) and a vector gauge potential $A_{\mu}=\left\{A_\varphi(r,x),0,0,A_t(r,x)\right\}$, compatible with the axisymmetric and stationary hypothesis, can be written as 
\bea 
     \label{ee-ernst-ch}  \left( \textsf{Re} \ \Er + | \mathbf{\Phi} |^2 \right) \nabla^2 \Er   &=&   \left( \overrightarrow{\nabla} \Er + 2 \ \mathbf{\Phi^*} \overrightarrow{\nabla} \mathbf{\Phi} \right) \cdot \overrightarrow{\nabla} \Er   \quad ,       \\
     \label{em-ernst}   \left( \textsf{Re} \ \Er + | \mathbf{\Phi} |^2 \right) \nabla^2 \mathbf{\Phi}  &=& \left( \overrightarrow{\nabla} \Er + 2 \ \mathbf{\Phi^*} \overrightarrow{\nabla} \mathbf{\Phi} \right) \cdot \overrightarrow{\nabla} \mathbf{\Phi} \quad .
\eea
When the Maxwell field is switched on the complex electromagnetic and gravitational Ernst potentials, for the metric (\ref{LWP-metric}), are respectively defined by
\beq \label{def-Phi-Er} 
       \mathbf{\Phi} := A_\varphi + i \tilde{A}_t  \qquad , \qquad \qquad     \Er := f - \mathbf{\Phi} \mathbf{\Phi}^* + i h  \quad ,
\eeq
where $\tilde{A}_t$ and $h$ stem from
\bea
    \label{A-tilde-e} \overrightarrow{\nabla} \tilde{A}_t &:=&  f \r^{-1} \overrightarrow{e}_\varphi \times (\overrightarrow{\nabla} A_t + \omega  \overrightarrow{\nabla} A_\varphi ) \ \ , \\
    \label{h-e}    \overrightarrow{\nabla} h &:=& - f^2 \r^{-1} \overrightarrow{e}_\varphi \times \overrightarrow{\nabla} \omega - 2 \ \textsf{Im} (\mathbf{\Phi}^*\overrightarrow{\nabla} \mathbf{\Phi} )  \ \ .
\eea

\subsection{Accelerating Kerr-Newman black hole in swirling universe}

The above construction allows us to deal with an accelerating charged black hole as a seed, for instance let consider the electromagnetic generalisation of the metric treated in sec. \ref{sec:generation}, i.e. the rotating charged C-metric\footnote{The full Plebanski-Demianski metric could be also used as a seed, which in practice means considering also the NUT parameter, but it complicates unnecessary the algebra of the solution by introducing further singularities encoden in the Misner string.}. It can be written in terms of the LWP metric (\ref{LWP-metric}) in the very same way of the uncharged case, through eqs. (\ref{f0})-(\ref{rho0}), the electric and magnetic charges enter only in the $\hat{\Delta}$ functions which replace eqs. (\ref{Dr})-(\ref{Dx}) and we will denote with an hat to distinguish from the uncharged case:
\bea
    \hat{\Delta}_r(r) &=& (1-A^2r^2)(r^2-2mr+a^2+e^2+p^2) \quad , \\
    \hat{\Delta}_x(x) &=& (1-x^2) [1-2Amx+A^2x^2(a^2+e^2+p^2)] \ \ . 
\eea
The seed electromagnetic field is given by
\beq
     A_\mu = \left\{ -\frac{er+pax}{r^2+a^2 x^2}, 0 , 0, -\frac{era(1-x^2)+px(r^2+a^2)}{r^2+a^2 x^2} \right\} \ \ .
\eeq

From the definition (\ref{A-tilde-e}) we find that 
\beq
           \tilde{A}_t = ex+ \frac{a(aex-pr)(1-x^2)}{r^2+a^2x^2} \ ,
\eeq
therefore the complex electromagnetic Ernst potential, as defined in (\ref{def-Phi-Er}) for the seed solution is 
\beq
    \hat{\mathbf{\Phi}}_0 (r,x) = \frac{(e+ip)(ia+rx)}{-ir+ax} \ .
\eeq
To obtain the gravitational Ernst potential $\hat{\Er}_0$, as in eq (\ref{def-Phi-Er}), for the seed metric, we only need to infer $\hat{h}_0(r,x)$, which, thanks to the definition (\ref{h-e}) is
\beq
            \hat{h}_0(r,x) = 2 a \left\{ m \left(\frac{1}{x}+x \right)  + \frac{r(1-x^2)\left[(e^2+p^2)(Ar-x)x - m(r-a^2A)(1-x^2) \right]}{x(1-Arx)(r^2+a^2x^2)}\right\}  \ .
\eeq
In order to embed the accelerating Kerr-Newman black hole into an external rotating background we apply the Ehlers transformation
\beq \label{Ernst-pot1}
\hat{\Er}_0 \longrightarrow \hat{\Er} = \frac{\hat{\Er}_0}{1+i\jmath\hat{\Er}_0} \qquad \quad ,  \quad  \qquad  \hat{\mathbf{\Phi}}_0 \longrightarrow  \hat{\mathbf{\Phi}} = \frac{\hat{\mathbf{\Phi}}_0}{1+i \jmath \hat{\Er}_0} \quad .
\eeq
These Ernst complex potentials uniquely identify the solution, then thanks to the definitions (\ref{A-tilde-e})-(\ref{h-e}) we can retrieve $\omega$ and $A_t$ to express the generated new solution in a standard form, as a metric and a vector potential. We recall that the $\gamma$ function remains unvaried, under the Ehlers transformation, with respect to the seed, also in the charged case.  \\
The final full expression for the metric and electromagnetic potential could be quite lengthy, we leave it in a Mathematica worksheet between the arXiv sources files. However it is worth to analyse possible conicity of the general case on the poles. The two condition for the accelerating Kerr-Newman metric\footnote{The intrinsic magnetic monopole charge of the black hole seed, $p$, is turned off for simplicity, nevertheless the resulting solution has magnetic charge provided by the intersection between the black hole electric field and rotation of the external gravitational background.} in the swirling universe are 
\beq \label{cony-e-1}
      \lim_{x \to -1} \int_0^{2\pi} \frac{d\varphi}{1-x^2} \sqrt{\frac{g_{\varphi\varphi}}{g_{xx}}} = 2\pi \Delta_\varphi \frac{1 + 2mA +A^2(a^2+e^2)}{1+8a \jmath m + \jmath^2(e^4+16a^2m^2)} \ \ ,
\eeq
and 
\beq \label{cony-e+1}
      \lim_{x \to +1} \int_0^{2\pi} \frac{d\varphi}{1-x^2} \sqrt{\frac{g_{\varphi\varphi}}{g_{xx}}} = 2\pi \Delta_\varphi \frac{1 - 2mA + A^2(a^2+e^2)}{1 - 8a \jmath m + \jmath^2(e^4+16a^2m^2)}   \ \ .
\eeq
From the above expressions it is clear that both conical singularities can be removed from the spacetime by properly set the range for the azimuthal angle, or by fixing $\Delta_\varphi$, and constraining one of the integrating constant of the solution, for instance $\jmath$, as done in the uncharged case:
\bea
      \jmath &=& \frac{2a+2a^3A^2+2aAe^2 \pm \sqrt{4a^2[1+A^2(a^2+e^2)]^2-A^2(e^4+16a^2m^2)}}{A(e^4+16a^2m^2)}  \ \ ,\\
      \D_\varphi &=& 4a \frac{2a+2a^3A^2+2aAe^2 \pm \sqrt{4a^2[1+A^2(a^2+e^2)]^2-A^2(e^4+16a^2m^2)}}{A^2(e^4+16a^2m^2)} \ \ .
\eea
Clearly these are the charged generalisation of the eqs. (\ref{c})-(\ref{Deltaphi}). Thanks to the above constraints the metric is completely free of conical singularities also in the presence of the Maxwell field. However note that the contribution of the electromagnetic charge is physically not relevant in the regularisation. Indeed again the spin-spin interaction between the angular momentum of the black hole and the rotation of the background is the fundamental justification for the black hole acceleration in the regular context. In fact if we have no intrinsic angular momentum in the black hole, i.e. $a=0$, the regularisation of the space-time cannot be pursued and one has to demand the existence of cosmic strings or struts in order to comprise the acceleration in the physical picture. 
Below we will show this feature in more detail working out explicitly the metric of the (accelerating) Reissner-Nordstrom black hole embedded into the swirling universe.

\subsection{Accelerating Reissner-Nordstrom black hole in the swirling universe} 

When the angular momentum of the seed black hole is switched off, that is $a=0$, the transformed Ernst potentials (\ref{Ernst-pot1}) take the form
\bea
      \hat{\Er}(r,x) &=& \frac{-(e^2+p^2)x^2-\frac{r^2(1-x^2)[1+Ax(-2m+Ax(e^2+p^2))]}{(1-Arx)^2}}{1+i\jmath \left[-(e^2+p^2)x^2-\frac{r^2(1-x^2)\{1+Ax[-2m+Ax(e^2+p^2)]\}}{(1-Arx)^2} \right]}  \ , \\
      \hat{\mathbf{\Phi}}(r,x) &=&  \frac{i(e+ip)x}{1-i\jmath \left[(e^2+p^2)x^2+\frac{r^2(1-x^2)\{1+Ax[-2m+Ax(e^2+p^2)]\}}{(1-Arx)^2} \right]} \ .
\eea 
From the definitions (\ref{def-Phi-Er})-(\ref{h-e}) we infer the metric and the electromagnetic vector potential. The metric can still be written as in eq. (\ref{swirling-sol}) but with a charged version of $\L(x,y)$ that we label with an hat: $\hat{\L}(r,x):=1+i\jmath \hat{\Er}_0=1+i\jmath[\hat{f}_0(r,x)-\hat{\mathbf{\Phi}}_0(r,x)] - \jmath \hat{h}_0(r,x)$ and 
\beq \label{omega-hat-RN-swirling}
      \hat{\omega}(r,x) = - \frac{2\jmath(e^2+p^2-2mr)}{Ar^2} +\frac{2\jmath(1-A^2r^2)(e^2+p^2+r^2-2mr)}{Ar^2(1-Arx)^2} + \hat{\omega}_0 \ .
\eeq
while the electric and magnetic components of the Maxwell potential are\footnote{A Mathematica worksheet with this dyonically charged C-metric in the swirling universe can also be found among the arXiv source files, for the reader's convenience.} respectively
\bea
     A_t &=& \frac{ \jmath p r(r-2m) + \jmath e^2 p (1-A^2r^2)x^2+e(1-Arx)^2 + \jmath px\{p^2x-r[2Ar(r-2m)+2mx-(1-A^2p^2)rx]\} }{-r(1-Arx)^2}     \nn      \\
       &-& \hat{\omega}(r,x)A_\varphi(r,x) \ . \nn \\
     A_\varphi &=& \frac{x(1-Arx)^2 \Big[p+\jmath e r^2-2Arx(p+e\jmath m r)+\{ A^2pr^2+\jmath e [q^2+r^2(A^2q^2-1)] \} x^2-2A\jmath e r (q^2-mr)x^2 \Big]}{-(1-Arx)^4-\jmath^2\big\{ q^2 x^2-2Arx^3q^2 + r^2[1+x(-x+A^2q^2-2mA(1-x^2))] \big\}^2} ,  \nn 
\eea
where $q:=\sqrt{e^2+p^2}$.\\
If we want to have a spacetime free from conical singularities we have to require that both
\beq \label{cony-ep-1}
      \lim_{x \to -1} \int_0^{2\pi} \frac{d\varphi}{1-x^2} \sqrt{\frac{g_{\varphi\varphi}}{g_{xx}}} = 2\pi \Delta_\varphi \frac{1 + 2mA +A^2(e^2+p^2)}{1 + \jmath^2(e^2+p^2)^2} \ \ 
\eeq
and 
\beq \label{cony-ep+1}
      \lim_{x \to +1} \int_0^{2\pi} \frac{d\varphi}{1-x^2} \sqrt{\frac{g_{\varphi\varphi}}{g_{xx}}} = 2\pi \Delta_\varphi \frac{1 - 2mA + A^2(e^2+p^2)}{1 + \jmath^2(e^2+p^2)^2}   \ \ 
\eeq
correspond to $2\pi$. Clearly this cannot be possible unless the mass or the acceleration parameters are null, which means respectively to remove the black hole or the acceleration from the charged C-metric in the rotational background. \\
The first possibility, that is $m=0$, ensures freedom from angular singularities but clearly it can not be interpreted as black hole, because the horizon becomes hyper-extremal, that is it disappears. It is not surprising the absence of conicity in this case, since even for $\jmath=0$ this monopoles presents no conical singularities. Of course the geometry displays naked curvature singularities at $r=0$, with or without $\jmath$, as the scalar curvature invariants, such as the Kretschmann, certifies. Therefore in this case we remain with an accelerating electric or magnetic monopole (in a swirling universe).\\
On the other hand in the latter case, i.e. $A=0$, the resulting Reissner-Nordstrom metric in the swirling background is completely regular outside the event horizon, hence it represents a legitimate black hole. The $A\to0$ limit is well behaved whenever the additive constant of the $\omega(r,x)$ function in (\ref{omega-hat-RN-swirling}) is fixed to absorb the divergent term, such as $\hat{w}_0=-2\jmath/A$. The explicit expression of the solution is given in appendix \ref{app:RN-swirling}. Actually it is the charged generalisation of the Schwarzschild metric in the external rotating background presented in \cite{swirling}. Note however that the solution is not static any more, after the Ehlers transformation the frame dragging of the external rotational background acts on the Reissner-Nordstrom black hole, which becomes stationary rotating too. Of course this is not in contrast with the unicity theorems in general relativity which states that the only rotating and charged black hole solution is Kerr-Newman, since the assumption of asymptotic flatness these theorems rely on is not accomplished.  \\
From the physical point of view it is clear why this happens: in order to have a regular spacetime one needs the simultaneous presence (or absence) of acceleration and spin-spin interaction, so that a balance can be achieved. But if we have just one of these two features, which brings in conical defects, it is not possible to ascribe the acceleration to the presence of the spin-spin interaction. \\
In the presence of the Electromagnetic field the correspondence between the extreme near horizon geometry of the black hole and the conformal field theory is realised, as done in section \ref{sec:cft-thermo}. Actually it is possible to take advantage also of the Maxwell U(1) symmetry to implement an alternative duality without considering the rotational degrees of freedom to reproduce the value of the entropy, as done in \cite{magnetised-RN-cft} or \cite{c-cft}.\\

\section{Summary and Conclusions}

In this article we exploit of the {\it ``magnetic Ehlers''} transformation, presented in \cite{swirling}, to remove the conical singularity of the rotating C-metric. We have shown that the interaction between the rotating background and the angular momentum of the Kerr black hole can interact to balance the angular defects naturally present in accelerating metrics. Hence we get a new regular black hole metric in pure general relativity. Physically it means that the black hole acceleration is due to the spin-spin interaction generated by the superposition of the two vacuum stationary and axisymmetric solutions.  \\
The tools of the gauge gravity duality are applied to the generated new solution in order to reproduce the microscopic entropy of the dual conformal field theory associated to the gravitational solution, at extremality. It coincides with the standard classical Bekenstein-Hawking entropy: a quarter of the event horizon area.\\    
The rigidity of the extreme near-horizon geometry, in analogy to the Meissner effect for black holes, let us suppose that a sort of Meissner effect can manifest also for the rotational external gravitational field, at zero temperature.\\
Then we extend the application of such Ehlers transformation to the Einstein-Maxwell theory. This allows us to build novel regular charged black hole space-times embedded into an external swirling universe. The more general solution we construct is the accelerating Kerr-Newman metric immersed into the rotating external gravitational field. However we infer that in general the black hole angular momentum is the key feature to get a smooth manifold, at least outside the event horizon. In fact we have explicitly worked out the metric for the dyonically charged C-metric in the swirling environment, but the only regular black hole subcase in this background was the non-accelerating charged one. The external gravitational field frame dragging adds rotation to the Reissner-Nordstrom black hole.\\
The procedure to remove the conical singularity from the C-metric resembles the one pioneered by Ernst with the help of the external electromagnetic Melvin universe \cite{ernst-remove}.  
Therefore, when our results are applied to the rotating C-metric, they open the possibility to discuss the pair creation rate of a couple of accelerating Kerr black holes at expense of the swirling external field, analogously to \cite{strominger}, \cite{hawking}, \cite{marcoa-pair}.   \\
The method to remove the conical singularity between the superposition of one or more stationary solutions to the swirling background introduced by the Ehlers transformation is general, it works in a seamless way, and does not apply only to accelerating metrics. Probably one of the more interesting applications for it could be the removal of the strut (or the strings) which maintain separated and at equilibrium the double Kerr (and Kerr-Newman) metric \cite{KramerNeug}. This application would generate one of the few  regular multi black hole rotating solutions in Einstein gravity\footnote{The equilibrium configuration between two Kerr Black holes can be reached also by using external gravitational background, such as the one proposed in \cite{charged-binary} or \cite{bubble}, but the latter has not been concretely realised yet.}. \\
From a theoretical perspective the main goal of this work consists in verify that within the standard model for gravitation, i.e. Einstein general relativity, the spin-spin interaction can provide a consistent source for black holes acceleration, without the contribution of any extra elements, at least when the collapsed star is rotating. \\
From a phenomenological perspective the existence of such swirling backgrounds, seems quite plausible in the context of galaxies merging, in particular for describing a stationary phase of the approaching process, when these objects are counter-rotating one with respect to the other. For modelling the full collision a time dependent solution would be necessary, but it is outside the domain of the solution generating technique at the moment. Possibly one has to apply to perturbation theory. However the physical phenomenon of the spin-spin intersection is expected to play a relevant role in that case as well.  \\
More generally the gravitational spin-spin interaction might take place in wider context such as spinning particles in a strong rotational environment, such as the accretion disk of a black hole where particles with spin can feel an axial force when their orbit get closer to the rotating collapsed stars. This effect could be of some relevance in the axial astrophysical or relativistic jets emitted by black holes, neutron stars or pulsar. \\   
Moreover it would be interesting to test the effect of the spin-spin interaction in system analogue to gravity, such as fluids. In fact, in that case, the force due to the rotational frame dragging interaction between two axisymmetric objects could be measurable in the laboratory.
\\
All these results can be directly extended to other theories which include minimally or conformally coupled scalar fields to general relativity, and also some scalar tensor theories \cite{marcoa-embedding},\cite{marcoa-stationary}. This is because the full symmetry group of the Ernst equations is preserved also in those modifications of the standard gravity model. It means that the procedure to embed a given seed metric into  the whirlpool background, described in \cite{swirling} and here, can be done for black holes, wormholes or any axisymmetric and stationary seed of these theories too, with or without Maxwell electromagnetic potential. Also the dilatonic coupling of the Maxwell field to general relativity is expected to work at least for some peculiar values of the coupling constant. Unfortunately in the presence of the cosmological constant the procedure described in this article cannot be pursued, because, the cosmological constant is known to break some relevant Lie point symmetries of the Ernst equations \cite{asto-lambda}.\\

\section*{Acknowledgements}
{\small We would like to thank Riccardo Martelli and Adriano Vigan\'o for fruitful discussions on the subject. %This work has been partially funded by INFN and by MIUR-PRIN contract 2017CC72MK-003} %n$^\textrm{o}$
\\

\appendix

\section{Reissner-Nordstrom black hole in the swirling universe}
\label{app:RN-swirling}

It is worth write explicitly one of the simpler and well behaved charged black hole solutions, the Reissner-Nordstrom spacetime embedded in the swirling universe. The metric can be written as in (\ref{swirling-sol}) 
\beq \label{swirling-f}
        ds^2 = -f(r,x) \Big( \Delta_\varphi d\varphi - \omega(r,x) dt \Big)^2 + \frac{1}{f(r,x)} \left[ \rho^2(r,x) dt^2 - e^{2\gamma_0(r,x)}  \left( \frac{{d r}^2}{\Delta_r(r)} + \frac{{d x}^2}{\Delta_x(x)} \right) \right] \ \ ,
\eeq
where
\bea
         f(r,x) &=& \frac{-r^2(1-x^2)}{1+\jmath^2[r^2+(e^2+p^2-r^2)x^2]^2} \ ,\\
         \omega(r,x) &=& \frac{4\jmath x\D_r(r)}	{r} + \omega_0 \ ,\\
         \gamma(r,x) &=& \frac{1}{2} \log \left[ r^4 \D_x(x) \right] \ ,\\
         \rho(r,x)  &=&  \sqrt{\D_r(r) \D_x(x)} \ , \\
           \D_r(r)   &=&  r^2-2mr+e^2+p^2  \\
           \D_x(x)   &=&  (1-x^2)           
\eea
While the no zero component of the electromagnetic potential are
\bea
    A_t(r,x) &=&  - \frac{e+\jmath p r(r-2m)+\jmath p \D_r(r)x^2}{r} -\omega(r,x) \frac{A_\varphi(r,x)}{\D_\varphi}  \ \ , \\
    A_\varphi(r,x) &=& - \frac{x[p+\jmath er^2 + \jmath e x^2(e^2+p^2-r^2)]}{1+\jmath^2[r^2+(e^2+p^2-r^2)x^2]^2} \D_\varphi \ \ .
\eea
No conical singularities are guaranteed on the axis of symmetry by choosing the constant $\D_\varphi = 1 + \jmath^2 (e^2+p^2)^2$.

\newpage

\section{$\omega$ for accelerating Kerr in swirling background}
\label{app-w}

It is convenient to express the $\omega(r,x)$ function as a finite power series expansion in terms of the rotational parameter $\jmath$
\beq
          \omega(r,x) = \frac{\omega_0 + \omega_1 \jmath + \omega_2 \jmath^2}{(a^2+r^2)^2(1-x^2)^{-1} \Delta_x - a^2 \Delta_r (1-x^2)} \ ,
\eeq
where 
\bea
         \omega_0 &=& 2m[Ax(a^2+r^2)-r(1-A^2r^2)]-aA^2(a^2+r^2) (r^2+a^2x^2) \nn \ ,\\
         \omega_1 &=& - 2 (1+a^2A^2)(r^2-2mr+a^2)\Big\{ -a^4Ax(1-2Arx+x^2)+r^3(1-2Amx)[Ar(1+x^2)-2x]   \nn \\
    &+& a^2\big[ 2A^2 r^3 x + 2x \big(m + (m-r)x^2\big) -Ar\big(4mx^2+r(1-x^4) \big) \big]  \Big\}/(1-Arx)^2  \nn  \ , \\
                   \omega_2 &=&  \Big\{  (1+4A^2m^2)r^4(1-2Amx)(1-Arx)^3   \\ 
                   &+& 2a^{10}A^7mx \big[1+6x^2+x^4-4Ar(x+3x^3)-A^2r^2(1-6x^2+3x^4) \big] \nn \\
                   &+& a^6 A^2 \Big[ 2x^2- 2Ax (m+3rx^2) + 2A^7 m r^5x\big(-8m+3r+6(r-3m)x^2 +(6m-r)x^4\big) \nn \\
                   &-& A^5r^2x\big( r^3x^2 (1+x^2) +8m^3(4+3x^2) +4m^2r(-2-23x^2+x^4)+2mr^2(1+6x^2+x^4)\big) \nn \\
                   &-& A^3x\big( -24m^3 + 4m^2r(-3+x^4) + r^3(3+3x^2+2x^4) + 2mr^2(-10+11x^2+2x^4)  \big) \nn \\
                   &+& 4A^6 mr^3 \big(2m^2 (-1+6x^2) + mr(4+11x^2-17x^4) +r^2(-1-4x^2+5x^4)  \big) \nn \\
                   &+& A^2 \big( -4m^2(1+4x^2+x^4) + 2mr(-1+9x^2+3x^4) +r^2(1+x^2+6x^4) \big)  \nn \\
                   &+& A^4r \big( 3r^3x^2(1+x^2)-2mr^2(1+6x^2)+8m^3(1-3x^2+x^4) + 4m^2r(-2-17x^2+11x^4) \big) \Big] \nn \\
                   &+& a^8 A^4 \Big[ x^2+4A^5mr^3x\big( m+r+6(r-m)x^2+rx^4\big) \nn \\ 
                   &+&A^3rx \big( -r^2x^4-4m^2(-3-15x^2+x^4) -2mr(9-2x^2+x^4) \big) \nn \\
                   &-&A^4mr^2 \big( m(4x^2-2) + 3r(1+6x^2+x^4) \big) \nn \\
                   &+& A^2 \big( 3r^2x^4-4m^2(1+9x^2+x^4) -2mr(1-2x^2+9x^4) \big) + A\big( 2m(x+6x^3+x^5-3rx^3)\big) \Big] \nn\\
                   &+& a^2r^2 \Big[ -r(Arx-1)^3(1+2A^2r^2+x^2)+8A^2m^3\big(1-x^2+A^6r^6x^4+A^3r^3x(5-4x^2) \nn \\
                   &+& A^4r^4x^2(x^2-12)+Arx(3x^2-5)+A^5r^5x(1+4x^2) +A^2r^2(-1+10x^2-3x^4) \big) \nn \\
                   &+& 2m \big(1-x^2+Arx(3x^2-5)+A^2r^2(-1+10x^2-3x^4)+A^3r^3x(2-10x^2-4x^4) \nn \\
                   &+& A^5r^5x(12+6x^2+x^4) +A^4r^4(3+6x^2+4^4) \big) - A^2m^2r \big[ -1-x^2+3A^3r^3x(x^2-1) +3Arx (1+x^2) \nn \\
                   &+& A^2r^2(1+3x^2-4x^4) + A^5r^5x (1+8x^2+x^4) +A^4r^4\big(1-3x^2(4+x^2) \big) \big] \Big] \nn \\
                   &+& a^4 \Big[ x^2-A(2mx+3rx^3) +A^2\big(2r (m+r) + 2(2m^2+2mr+r^2)x^2+3r^2x^4 \big)\nn \\
                   &+& 2A^8mr^5 \big[ r(r-2m) + 6(r-2m)(m+r)x^2+(-12m^2+6mr+r^2)x^4 \big] \nn \\
                   &+& 4A^9m^2r^6x\big( m(2+6x^2) -r(1+3x^2+x^4) \big) -A^3x\big(8m^3+10mr^2+12m^2rx^2+r^3(6+6x^2+x^4) \big) \nn \\
                   &+& A^7 r^4 x \big( -r^3x^2+48m^3(2x^2-1) -4m^2r(-14+16x^2+x^4) -2mr^2(9-3x^2+2x^4) \big)  \nn \\
                   &+& A^6r^3 \big(3r^3x^2-4m^2r(2-20x^2+x^4) +8m^3(2-8x^2+x^4)+2mr^2(1-5x^2+14x^4)\big) \nn \\
                   &+& A^5r^2x \big[ m^3(56-96x^2) - 4mr^2x^2(14+x^2) + 4 m^2 r (-14+14x^2+3x^4) -r^3 \big(3+2(x^2+x^4) \big) \big] \nn \\
                   &+& A^4 r \big[ 2mr^2(1+22x^2) + 8m^3(-2+8x^2+x^4) -4m^2r(-2+10x^2+x^4) + r^3\big( 1+6(x^2+x^4) \big) \big] \Big]           \Big\} \nn   \\         & \big/ & \Big\{ aA^4(-1+Arx)^3 \Big\}  \nn
\eea

\end{document}